\documentclass{article}

\input pz.sty
\input epsf.sty
\usepackage{longtable}

\begin{document}
\PZtitletl{$UBV$ Photometry of the Post-AGB Star} {IRAS
22272+5435=V354 Lac in 1990-2008}

\PZauth{V. P. Arkhipova, N. P. Ikonnikova, G. V. Komissarova}
\PZinsto{Sternberg Astronomical Institute, University Ave. 13,
119992 Moscow, Russia}

\begin{abstract}

New $UBV$-photometry obtained in 2000-2008 is presented for the
post-AGB star IRAS 22272+ 5435=V354 Lac. The star showed
semi-regular light variations with varying amplitudes. The maximal
amplitude did not exceed: $\Delta V$=0.$^{m}$5, $\Delta
B$=0.$^{m}$7 and $\Delta U$=1.$^{m}$0. For 2000-2008, we have
found a photometric period near 128 days. The analysis of
long-term observations in 1990-2008 reveals variations with two
close periods: 128 and 131 days, causing amplitude modulation. The
$V$-$(B-V)$ diagram shows a clear correlation: the star is
generally bluer when brighter. From our $UBV$ data, we derive
$E(B-V)$=0.5 and conclude that the spectral type of the star
varies between K1 to K7 during pulsations. The mean $UBV$-data of
V354 Lac have not changed during the past 19 years: $V=8.^{m}60$,
$B-V=2.^{m}06$ and $U-B=2.^{m}14$.

\end{abstract}
\bigskip
\bigskip
\PZsubtitle{INTRODUCTION}

The IR source IRAS~22272+5435 is identified with the bright star
BD+54$^{\circ}$2787=HD 235858 ($22^{h} 29^{m} 10^{s} +54^{\circ}
51' 06''$(2000)). It is one of the most reliable protoplanetary
nebula candidates.

The spectral energy distribution of IRAS~22272+5435 show a
characteristic double peak, with about the same amounts of energy
emitted in the visible plus near-infrared (from the reddened
photosphere) and in the mid-infrared (re-emission from
circumstellar dust) (Hrivnak and Kwok 1991).

The chemical composition of IRAS~22272+5435 is typical of carbon
post-AGB stars. Za\v{c}s, Klochkova and Panchuk (1995) found the
star to be iron poor: [Fe/H]=--0.49. The elements of the
$\alpha$-process are overabundant relative to the solar
compositions and the carbon abundance is C/O$\approx$12. In the
spectra of IRAS~22272+5435  strong molecular bands of C$_{3}$ and
C$_{2}$ are found (Hrivnak and Kwok 1991).

In the survey of protoplanetary nebulae candidates with Hubble
Space Telescope, Ueta et al. (2000) discovered an elongated
low-surface-brightness reflection nebulosity around IRAS
22272+5435. These authors believe that the multilobed nebula may
be a progenitor of a complex planetary nebula.

The spectral classification of V354 Lac is a difficult problem
because of its peculiar spectrum. In the HD catalog, HD 235858 has
the spectral type K5. McCuskey (1955) classified the object as
M0III and Hrivnak and Kwok (1991), as GpIa. The atmospheric
parameters of IRAS 22272+5435, $T_{eff}$=5600 K and $\log g$=0.5,
correspond to a G2 supergiant (Za$\check{c}$s, Klochkova and
Panchuk, 1995).

The brightness variability of BD+54$^{\circ}$2787 was discovered
by Strohmeier and Knigge (1960). They list it as a possible
short-period variable with an amplitude of 0.$^{m}$5. Filatov
(1961) reported irregular variability of the star with an
amplitude of 1.$^{m}$5-2$^{m}$.0. In the 62nd GCVS Name-List
(Kukarkin et al., 1977), BD+54$^{\circ}$2787 got the designation
V354~Lac. The GCVS variability currently listed for the star is LB
(a slow irregular variable).

The light-curve and radial-velocity studies of V354 Lac were
carried out by Hrivnak and Lu (2000) respectively in 1994-1996 and
1991-1995. They found brightness and radial-velocity variability
with the period $P$=127$^{d}$.

The radial-velocity monitoring of V354 Lac performed by Za\v{c}s
et al. (2009) confirmed regular variations with a peak-to-peak
amplitude of about 10 km s$^{-1}$ and a period of about 131.2
days.

\bigskip
\bigskip
\PZsubtitle{$UBV$ OBSERVATIONS OF V354~Lac}

Our $UBV$ photometry of IRAS~22272+5435 was performed in 1990-2008
with the 0.6-m telescope of the Crimean Station of the Sternberg
Astronomical Institute. The measurements were obtained using a
pulse-counting photometer with an EMI 9789 photomultiplier, and a
filter set consistent with the Johnson system. All observations
were made with respect of the comparison star BD+54$^{\circ}$2793
($V=8.^{m}$54, $B=10.^{m}$45, $U=12.^{m}$79). The typical
photometric uncertainties range from 0.$^{m}$01 in the $V$ band
and to 0.$^{m}$05 in the $U$-band. The observations obtained in
1990-1999 were published earlier (Arkhipova et al., 1993 and
Arkhipova et al., 2000).
The light and color curves of V354~Lac for
1990-2008 are presented in Fig.1.

\PZfig{10cm}{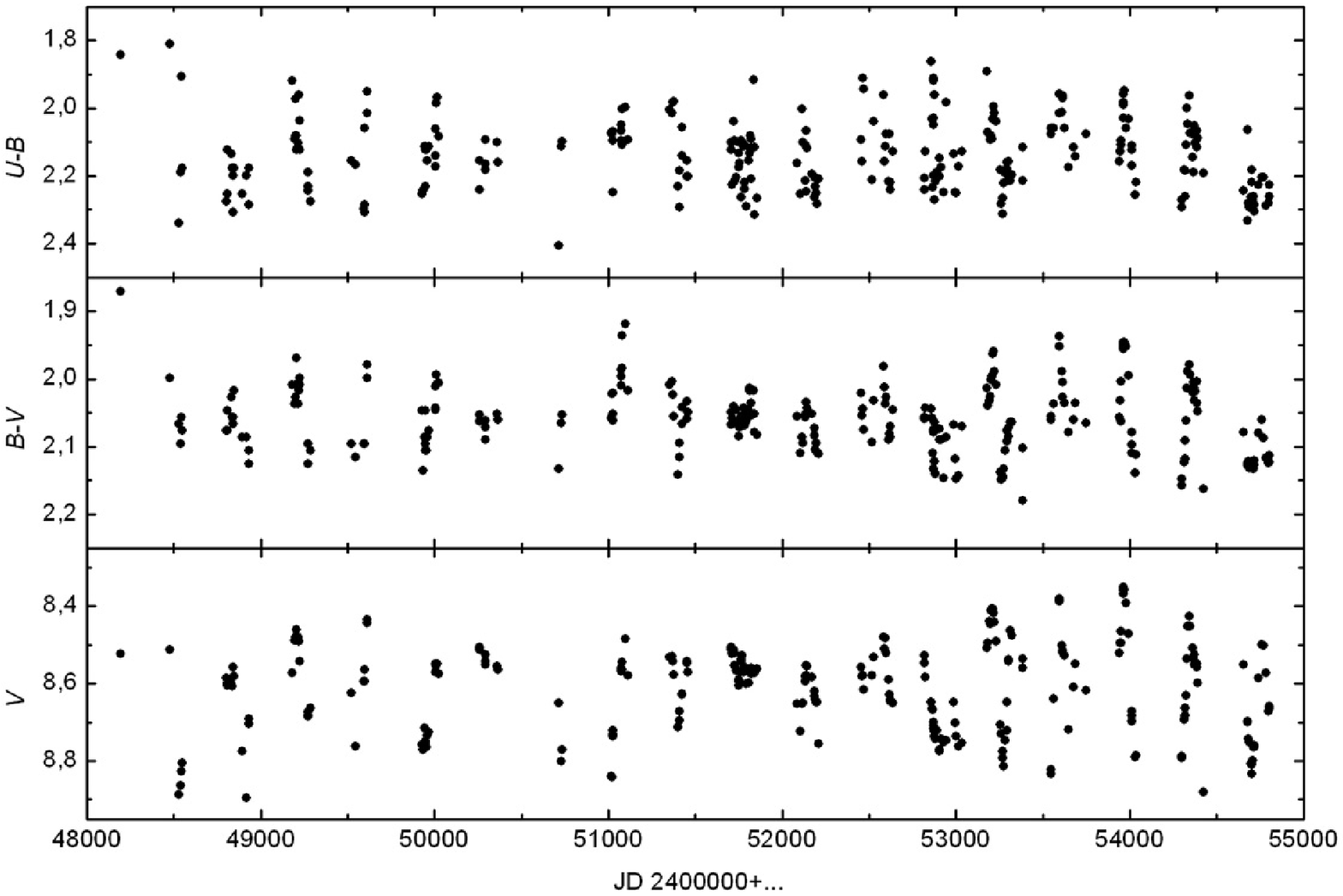}{The light and color curves of V354~Lac in
1990-2008.}

During our observations, V354~Lac showed semi-regular brightness
variations with varying amplitudes. Its largest amplitudes do not
exceed $\Delta V$=0.$^{m}$5, $\Delta B$=0.$^{m}$7, $\Delta
U$=1.$^{m}$0.

We searched for periodicity in the observations obtained in
1990-2008 using the DFT (Discrete Fourier Transforms) package by
Dr. V.M.Lyuty.

In the power spectrum (Fig.2), triplet frequencies are dominating:
$\nu_{1}$=0.00781, $\nu_{2}$=0.00763, $\nu_{3}$=0.00800. The
ratios $\nu_{1}$/$\nu_{2}$ and $\nu_{3}$/$\nu_{1}$ are close to
1.024. The periods $P_{1}$=1/$\nu_{1}$=128$^{d}$ and
$P_{2}$=1/$\nu_{2}$=131$^{d}$ have approximately the same
amplitudes, about 0.$^{m}$18 in the $V$ band.

\PZfig{5cm}{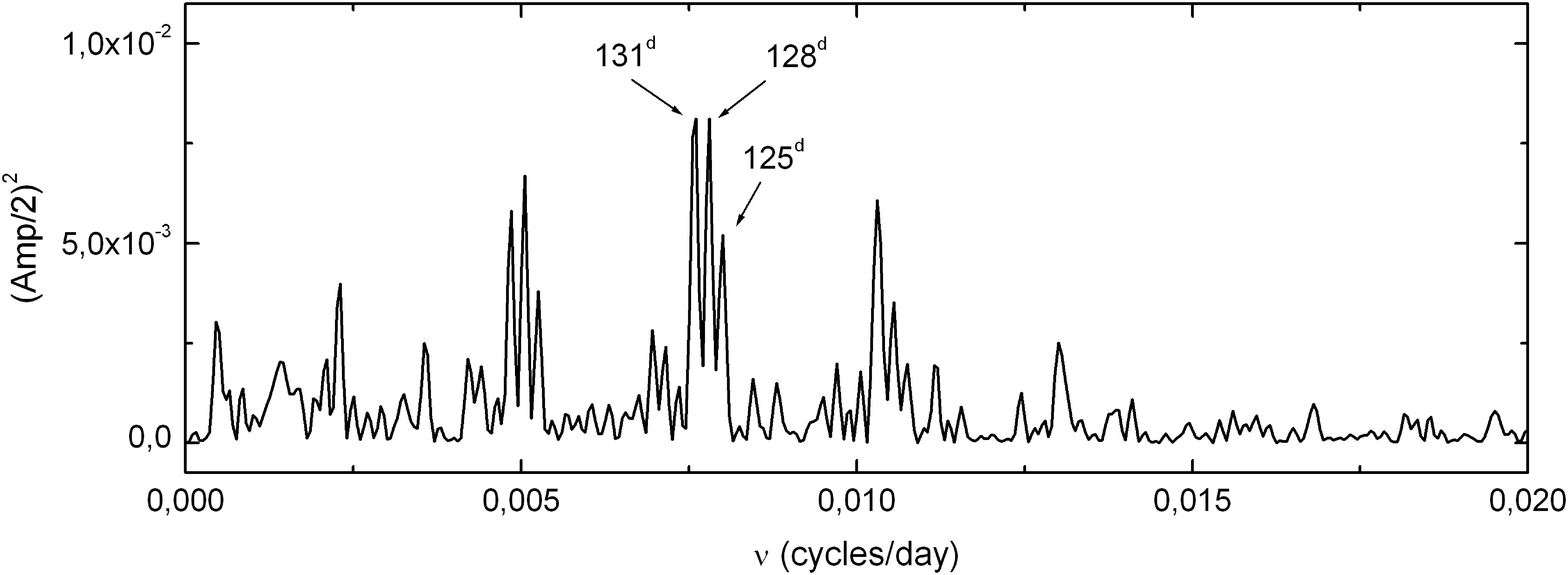}{The power spectrum of V354~Lac for
observations of 1990-2008.}

A single period, $P=128\pm 2$ days, was found in the observations
of 2000-2008. The corresponding power spectra and the $V$-band
phased light curve are shown in Fig.3.

\PZfig{7cm}{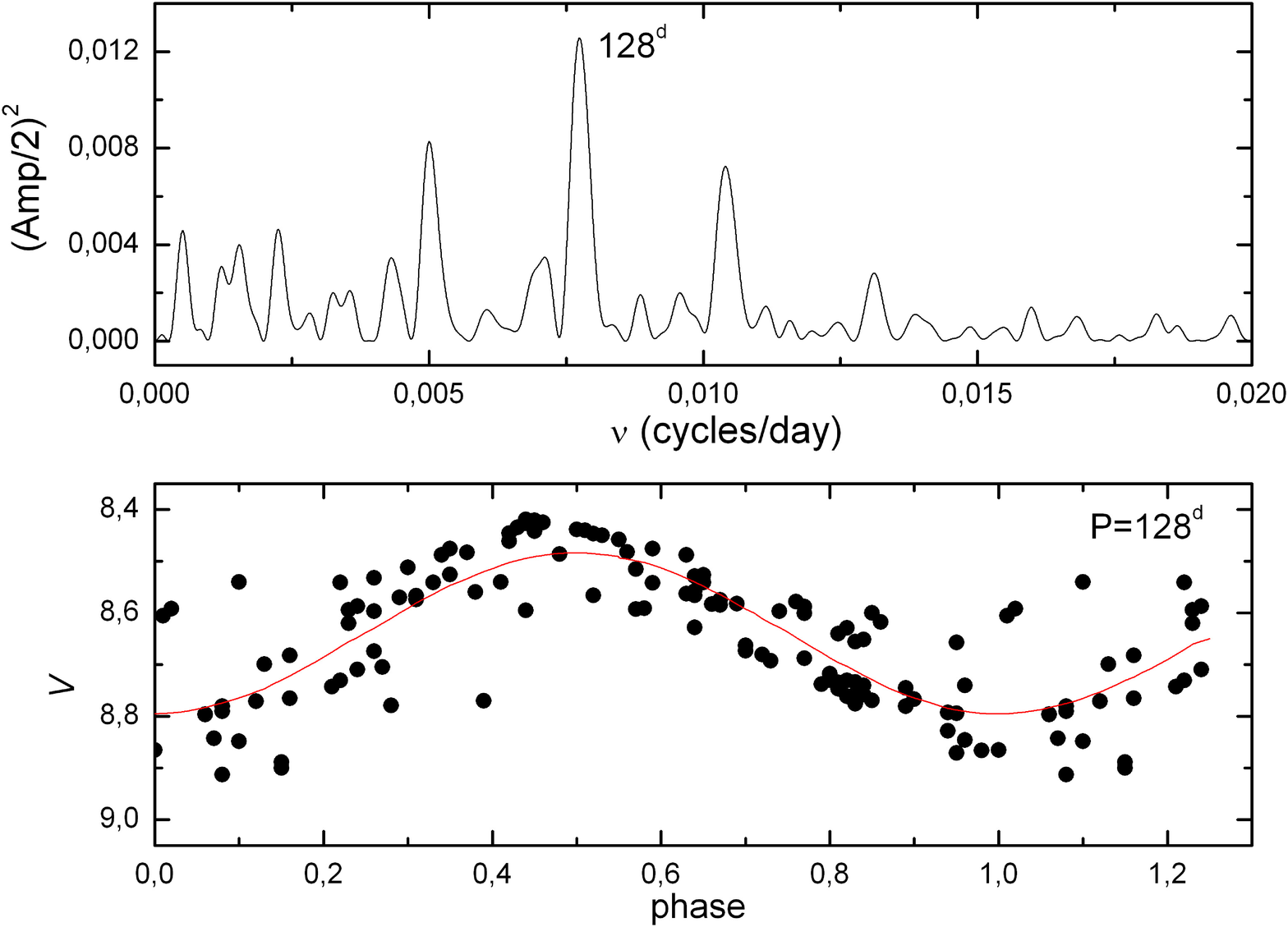}{The power spectrum (top panel) and the
phased $V$ light curve (bottom panel) of V354 Lac for observations
of 2000-2008.}

Considering the radial-velocity variations (Za\v{c}s et al., 2009)
together with our light curve (Fig.4) confirms the conclusion of
Hrivnak and Lu (2000) that V354 Lac "is brightest when it is at
its average size and expanding and faintest when at its average
size and contracting". Hrivnak and Lu interpreted this variability
as due to pulsation in the star rather than its binary nature.

\PZfig{4cm}{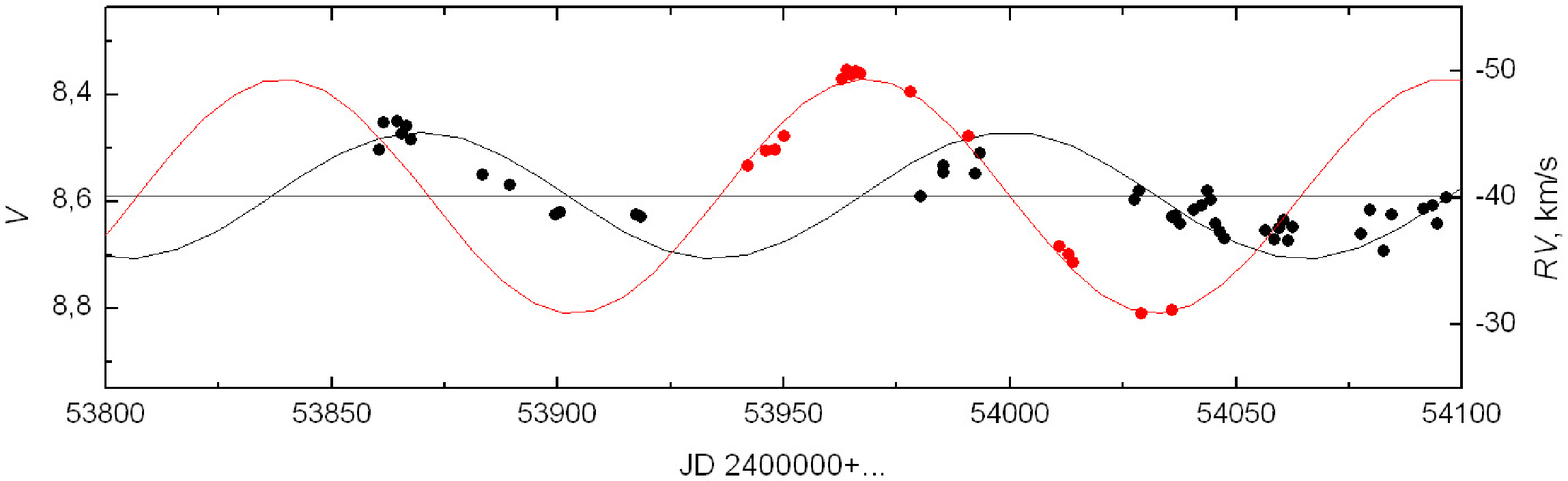}{The radial-velocity variations of V354~Lac
(black filled circles) with a sinusoid fit ($P$=131.2 days; black
line) and $V$ light curve (red filled circles) with a sinusoid fit
($P$=128 days; red line).}

The $V$-$(B-V)$ diagram shows a clear correlation: the star is
generally bluer when brighter (Fig.5).

\PZfig{6cm}{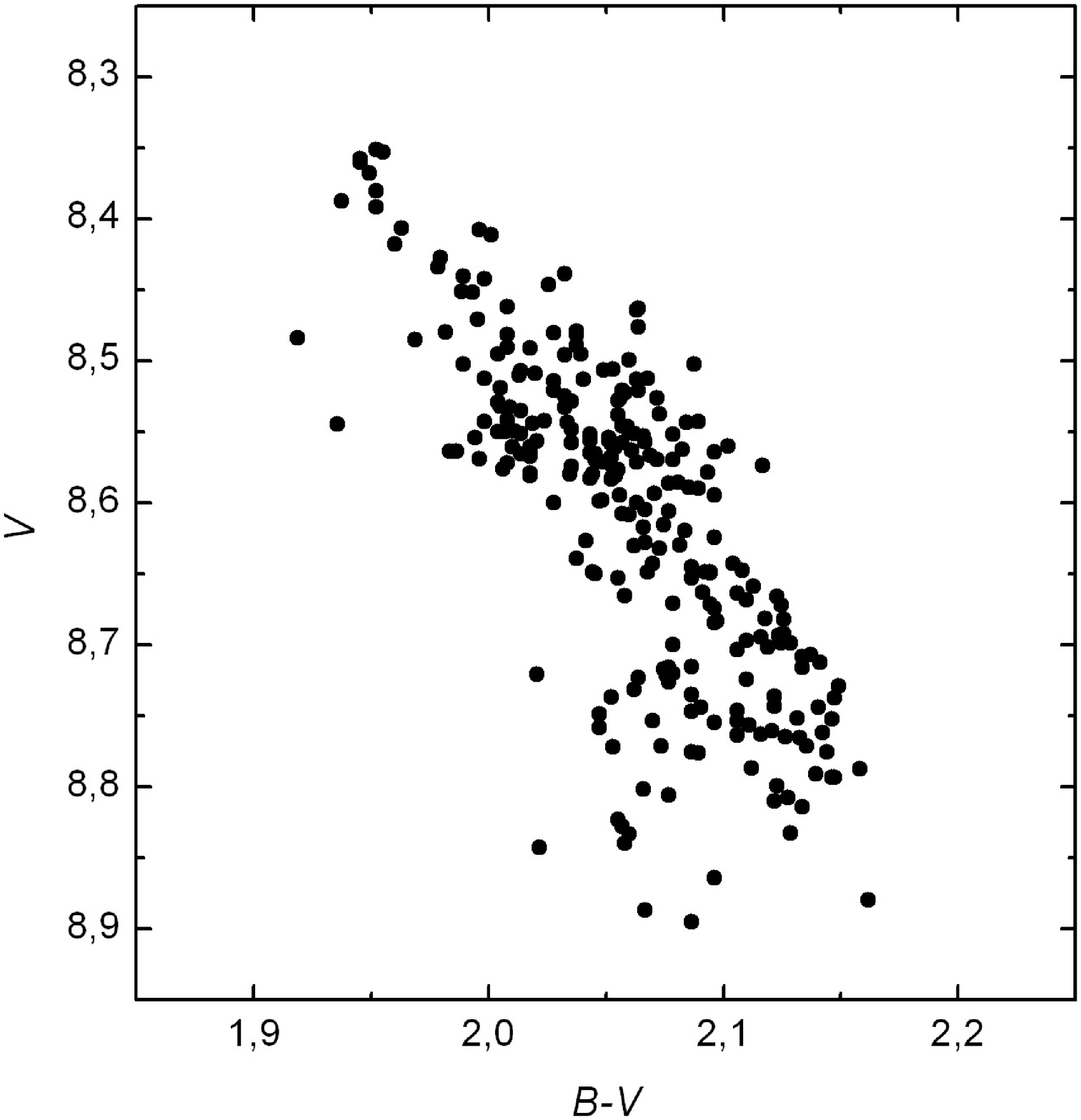}{The color-brightness diagram for V354~Lac.}

V354~Lac is located at a low galactic latitude
($b=-2.^{\circ}52$). The interstellar extinction in the direction
of V354 Lac from maps of Neckel and Klare (1980) is $1.^{m}2 <
A_{V} < 1.^{m}9$ at $r$=1 kpc. From $UBV$ data, we conclude that
$E(B-V)$=0.5 and $A_{V}=1.^{m}55$. Thus, practically all the color
excess can be caused by interstellar extinction.

\PZfig{7cm}{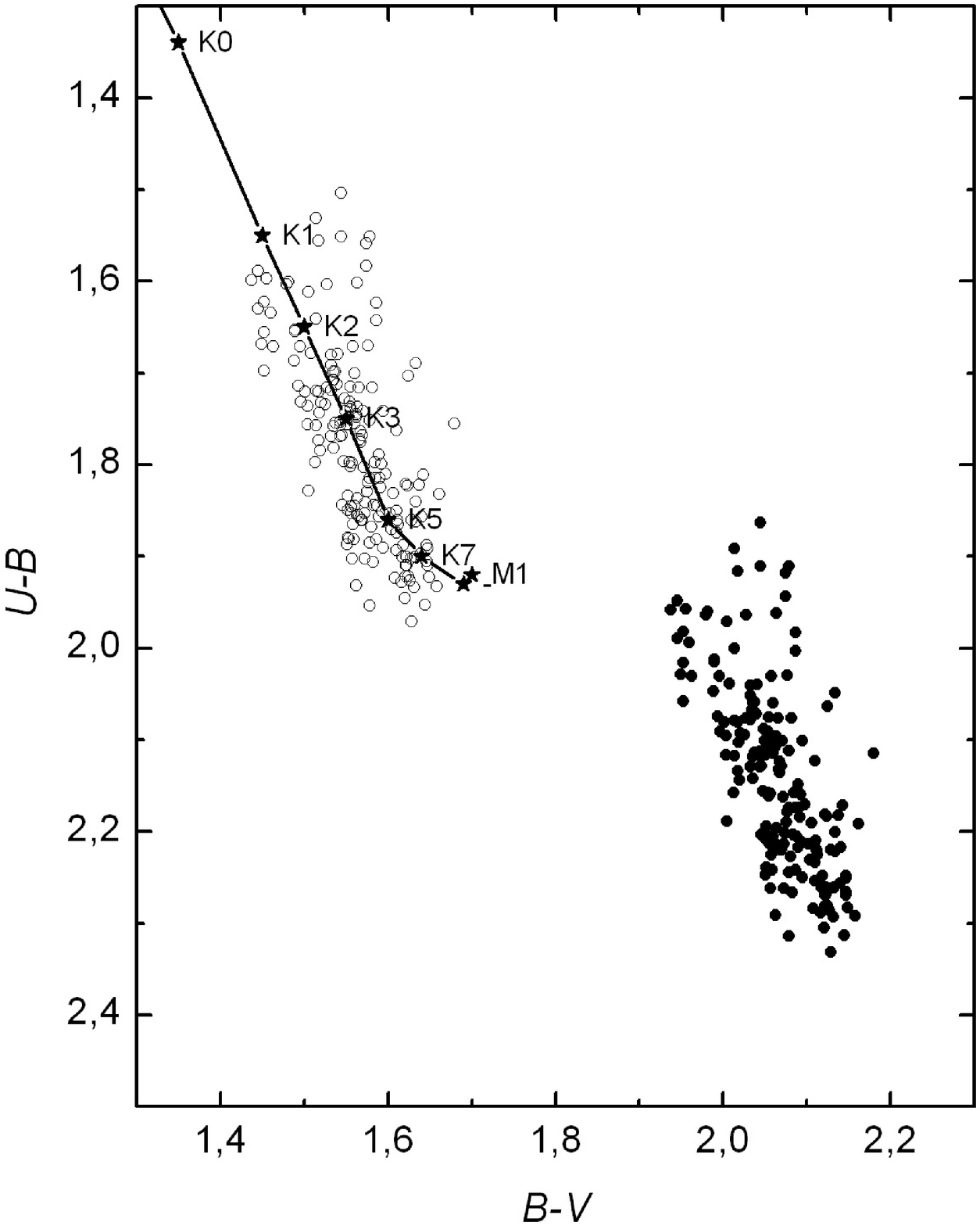}{V354 Lac in the   $(U-B)$-$(B-V)$ two-color
diagram. The solid curve is the supergiant sequence according to
Strai\v{c}ys (1977); the dots represent our observations and the
open circles represents de-reddened data with $E(B-V)=0.5$.}

In the $(U-B)$-$(B-V)$ two-color diagram (Fig.6), the star moves
along the sequence of supergiants in the course of its
fluctuations. Correction of the color indices for reddening with
$E(B-V)$=0.5 put the star on the sequence of supergiants, where
the spectral type of the star varies during pulsations from K1 to
K7.

\bigskip
\bigskip
\PZsubtitle{CONCLUSIONS}

Our long-term $UBV$ photometry of the post-AGB star V354 Lac
permitted to study its variability character. V354 Lac showed
semi-regular light variations with varying amplitudes. The maximal
amplitudes did not exceed: $\Delta V$=0.$^{m}$5, $\Delta
B$=0.$^{m}$7, $\Delta U$=1.$^{m}$0. The observed behavior of V354
Lac is explained with beating of two closely spaced pulsation
modes with the period ratio 1.02. The other post-AGB stars with
amplitude modulation are IRAS 19386+0155=V1648 Aql (Arkhipova et
al., 2009) and IRAS 08544-4431 (Kiss et al., 2007). The period
ratio for them is 1.04. Some pulsating red variables on the
asymptotic giant branch (AGB) show pulsations with two close
periods. Such a phenomenon is detected, for example, for RX UMa
and RY Leo (Kiss et al., 2000). The authors of the cited paper
assumed that the period ratios assumed that the period ratios
found for these stars (1.03-1.10) suggested either high-order
overtone or radial+non-radial oscillation.

The mean $UBV$ parameters of V354 Lac has not changed during the
past 19 years: $\overline{V}=8.^{m}60$,
$\overline{(B-V)}=2.^{m}06$ and $\overline{(U-B)}=2.^{m}14$.

The brightness and the colors change in phase: the star is
generally bluer when brighter. The temperature variations during
pulsations correspond to the changes of the spectral type from K1
to K7.

\medskip

{\bf Acknowledgements:} The study was partly supported by the
Council for the Program of Support for Leading Scientific Schools
(projects NSh.433.2008.2).

\references Arkhipova, V.P., Ikonnikova, N.P., Noskova, R.I.,
1993, {\it Astronomy Letters}, {\bf 19}, 169

Arkhipova, V.P., Ikonnikova, N.P., Noskova, R.I., et al., 2000,
{\it Astronomy Letters}, {\bf 26}, 609

Arkhipova, V.P., Ikonnikova, N.P., Komissarova, G.V., {\it
Astronomy Letters} 2010, {\bf 36} (in press) (arXiv:0911.0268)

Filatov, G.S., 1961, {\it Astron. Tsirk.}, No. 223

Hrivnak, B.J. and Kwok, S., 1991, {\it Astrophys. J.}, {\bf 371},
631

Hrivnak, B.J. and Lu, W.X., 2000, {\it The Carbon Star Phenomenon.
IAU Symp. 177} (Ed. R. F. Wing, Dordrecht: Kluwer Acad. Publ.), p.
293

Kiss, L.L., Szatm\'{a}ry, K., Szab\'{o}, G.M. and Mattei, J.A.,
2000, {\it Astron. Astrophys., Suppl. Ser.}, {\bf 145}, 283

Kiss, L.L., Derekas, A., Szab\'{o}, G.M. , et al., 2007, {\it
MNRAS}, {\bf 375}, 1338

Kukarkin, B.V., Kholopov, P.N., Fedorovich V.P., et al., 1977,
{\it Inform. Bull. Var. Stars}, No. 1248, 1

Neckel, Th. and Klare, G., 1980, {\it Astron. Astrophys., Suppl.
Ser.}, {\bf 42}, 251

Strai\v{c}ys, V. L., 1977,  {\it Multicolor Stellar Photometry}
(Mokslas: Vilnius)

Strohmeier, W., Knigge, R., 1960, {\it Ver\"{o}ff.
Remeis-Sternwarte Bamberg}, {\bf 5}, No. 5, 1

Ueta, T., Meixner, M. and Bobrowsky, M., 2000,  {\it Astrophys. J.},
{\bf 528}, 861

Za\v{c}s, L., Klochkova, V.G. and Panchuk, V.E., 1995, {\it
MNRAS}, {\bf 275}, 764

Za\v{c}s, L., Sperauskas, J., Musaev, F.A et al., 2009, {\it
Astrophys. J.}, {\bf 695}, 203

\endreferences
\end{document}